# Josephson Current in Rashba-based Superconducting Nanowires with Geometric Misalignment

Z.-J. Ying , M. Cuoco, P. Gentile
CNR-SPIN and Dipartimento di Fisica "E.R. Caianiello",
Università di Salerno
Fisciano (SA), 84084, Italy

C. Ortix
Institute for Theoretical Physics, Center for Extreme Matter
and Emergent Phenomena, Utrecht University,
Princetonplein 5, 3584 CC Utrecht, The Netherlands

*Abstract—* We investigate the properties of a weak link between two Rashba-based superconducting nanowires with geometric misalignment. By applying an external magnetic field the system can be driven into a topological non-trivial regime. We demonstrate that the Josephson current can be modulated in amplitude and sign through the variation of the applied field and, remarkably, via the angle controlling the spin-orbit locking mismatch at the interface of the nanowires. The proposed setup with misaligned coplanar nanowires provides the building block configuration for the manipulation of coherent transport via geometric-controlled mixing/splitting of interface states.

## I. Introduction

Low-dimensional semiconducting nanomaterials in the presence of inversion symmetry breaking play a relevant role in the area of spintronics, spin-orbitronics and topological states of matter. Apart from conventional material geometries the most recent advances in nanotechnology led to have at hand an entirely novel family of low-dimensional nanostructures: flexible semiconductor nanomaterials which are bent into curved, deformable objects ranging from semiconductor nanotubes, to nanohelices, etc. Indeed, the role of geometric [1] deformation in Rashba spin-orbit coupled nanostructures has been shown to drive metal-insulator transitions and non-trivial topological states of matter [2] as well as spin textures with a tunable winding [3], thus indicating a feasible path for an all-geometric-electric control of the electron spin interference in deformed quantum rings [3].

Inversion symmetry breaking is also a crucial physical ingredient for superconducting quantum systems especially for its potential to yield both spin manipulation of the Cooper pairs and non-trivial dissipationless charge-and-spin transport. For instance, the effect of the Rashba spin-orbit coupling (RSOC) on the Josephson current has been widely studied in the presence of a source of time reversal symmetry breaking, e.g., by a magnetic-field induced Zeeman splitting or by magnetic exchange interactions [4,5]. Remarkably, a mechanical and electric manipulation of the Cooper pair transmission has been theoretically proposed by employing a non-trivial geometric profile of a RSOC bent nanowire between two superconducting leads [6].

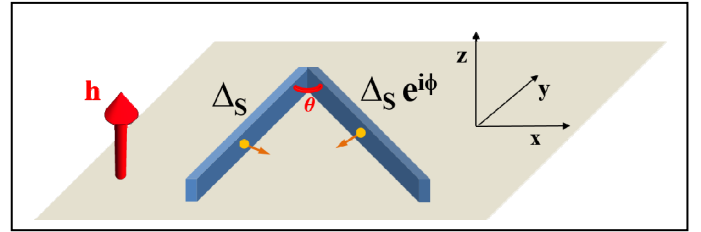

Fig. 1. Schematic representation of the superconducting nanowires with different orientations in the x-y plane such as to form an angle $\theta$. $\phi$ is the phase drop between the two spin-singlet superconductors and **h** is the applied magnetic field perpendicular to the plane where the nanowire are placed.

A major boost in the framework of inversion asymmetric systems relied on the proposal [7, 8] and the observation of a topological superconducting phase [9–12], hosting end Majorana modes, in a one-dimensional semiconductor nanowire with sizable RSCO and proximity-induced superconductivity.

Motivated by the excitement in both topological states of matter and novel shape deformed nanostructures, we have now theoretically considered the consequences of geometric effects on the coherent transport properties of superconducting RSOC nanowires. For such purpose, we employ a setup that encodes the basic element of geometric deformation by interfacing two nanowires with misaligned spatial orientations in the plane (Fig. 1). Such configuration is fundamental to explore the basic effects of curvature on the Josephson transport. The main resulting outcomes indicate that the Josephson current can be modulated in amplitude and sign through both the angle controlling the spin-orbit locking mismatch at the interface of the nanowires and the variation of the applied field.

## II. Model and Methodology

The model system is made of two nanowires of size L (in units of the lattice constant) extending in the *x-y* plane, and placed in such a way to form a relative angle $\theta$ (see Fig. 1). The electrons move along each one dimensional nanostructure in the presence of a local spin-singlet pairing interaction. Due to the structural inversion symmetry breaking the electrons are also subject to a RSOC, which couples the orbital and the local spin component normal to the electron motion. Since the two nanowires are misaligned and not collinear, the spin-orientation perpendicular to the momentum of the quasiparticle changes its direction when the electrons move through the nanowire crossing point at the interface. The Hamiltonian is then composed by the kinetic single particle term, the RSOC and the local superconducting pairing interaction. For convenience, we assume that the RSOC and pairing interaction have the same amplitude in the two nanowires. Moreover, we consider the application of an external magnetic field oriented perpendicular to the plane of the junction.

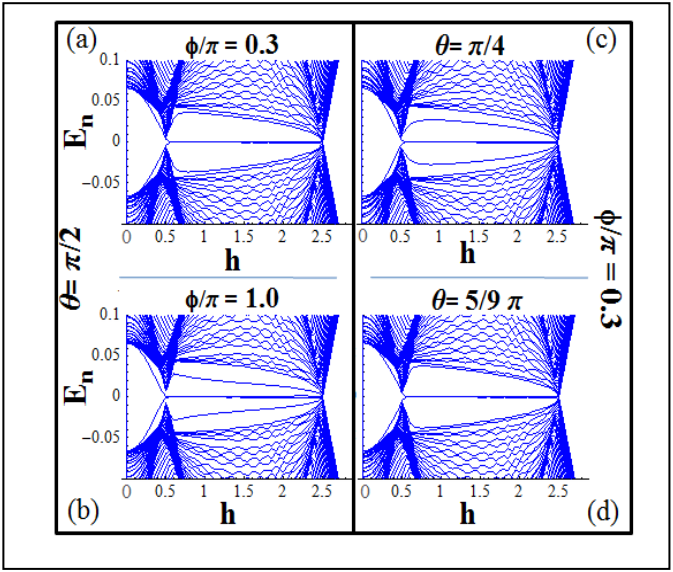

Fig. 2. Field dependence of the energy spectrum for a given orientation mismatch $\theta=\pi/2$ of the RSOC chains and two values of the superconducting phase difference $\phi=0.3\pi$ (a) and $\pi$ (b). Magnetic field evolution of the energy spectrum at a given $\phi=0.3\pi$ and for two different misalignment angles of the nanowires $\theta=\pi/4$ (c) and $\theta=5\pi/9$ (d). All the energies are in units of half the bandwidth $2t$ and the chemical potential is $\mu=1.5$.

Then, after decoupling the pairing term, we have to deal with the conventional Bogoliubov–de Gennes (BdG) type Hamiltonian with RSOC and local spin-singlet pairing. Then, we solve the BdG equations on a lattice and we determine the spin-resolved energy spectrum of the system, including both bulk and the edge Andreev states. We assume the pairing amplitude to be uniform along the nanowire. Furthermore, in order to investigate the effects of the phase difference between the two spin-singlet superconductors, we employ the conventional procedure for the analysis of the Josephson current, $J_S$, by transforming the order parameter $\Delta_S$ with the phase factor $\exp[i\phi]$ (see Fig. 1).

## III. RESULTS

We start the discussion by considering the analysis of the energy spectrum $E_n$ obtained from the numerical solution of the Bogoliubov–de Gennes equations. In Fig. 2 we show the magnetic field evolution of the energy spectrum for various representative values of the superconducting phase difference $\phi$ and orientation misalignment angle $\theta$ of the RSOC chains. At zero field the system has a gap that closes at $h\sim0.5$ and $h\sim2.5$, signaling the occurrence of topological phase transitions. Indeed, in this window of applied fields, one gets zero energy bound states corresponding to Majorana modes that are localized at the two edges which are far from the crossing points of the RSOC chains. On the other hand, the modes at the interface are at finite energy as they are fermionic bound states resulting from the hybridization of two Majorana modes. As expected, the interface in-gap modes are dispersing due to the phase difference $\phi$ (Figs. 2 (a),(b)). Interestingly, a variation of the geometric configuration of the two nanowires through the angle $\theta$ also affects the in-gap modes as explicitly demonstrated by comparing two different misalignment angles in Figs. 2(c),(d).

These features contribute significantly to the modification of the Josephson current especially when the system is in the topological non-trivial phase. We find that the usual oscillating behaviour is altered in the topological regime if compared to the non-topological configuration when analysing the configuration with $\theta=\pi/2$ (Fig. 3 (a)). Indeed, in the topological regime the Josephson current starts to be negative at $\phi=0$ and exhibits a sign change already for a phase drop of $\phi=\pi/2$. Remarkably, the misalignment angle $\theta$ can drive an amplitude modulation in the topological trivial state (Fig. 3 (b)), while in the region of applied field with in-gap bound states the misalignment angle can induce a sign change in the Josephson current. Such behavior is general and occurs at any phase drop between the two RSOC nanowire.

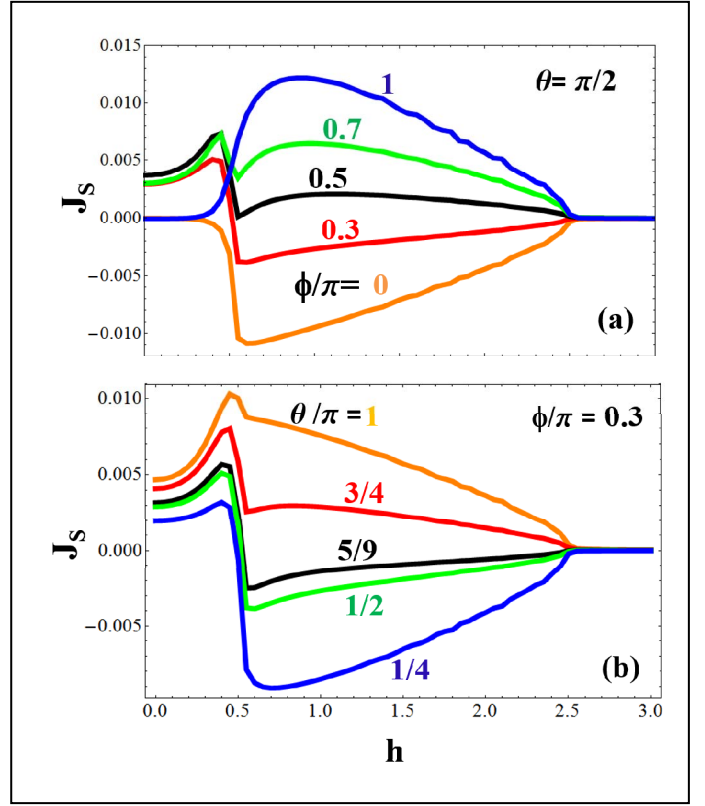

Fig. 3. Evolution of the Josephson current as a function of the applied magnetic field for different phase differences $\phi$ at a given misalignment angle $\theta=\pi/2$ (a) and for different values of $\theta$ for a representative $\phi=0.3\pi$ (b).

Finally, the observed behavior of the RSOC misaligned superconducting nanostructures clearly indicates the possibility of having a control of the interface spin mixing and of the Josephson current through a geometric parameter. Such results point to a wide area of novel design of Josephson junctions networks where the geometric configuration can be employed to design the overall transport properties.

ACKNOWLEDGMENT

We acknowledge the financial support of the Future and Emerging Technologies (FET) programme under FET-Open grant number: 618083 (CNTQC). C.O. acknowledges support from the Deutsche Forschungsgemeinschaft (grant No. OR 404/1-1) and from VIDI grant (project 680-47-543) financed by the Netherlands Organization for Scientific Research (NWO).

REFERENCES

[1] P. Gentile, M. Cuoco, C. Ortix, SPIN, Vol. 3, No. 2, 1340002 (2013).
[2] P. Gentile, M. Cuoco, C. Ortix, Phys. Rev. Lett. 115, 256801 (2015).
[3] Z.-J. Ying, P. Gentile, C. Ortix, and M. Cuoco, Phys. Rev. B 94, 081406(R) (2016).
[4] E. V. Bezuglyi, A. S. Rozhavsky, I. D. Vagner, and P.Wyder, Phys. Rev. B 66, 052508 (2002).
[5] A. A. Reynoso, Gonzalo Usaj, C. A. Balseiro, D. Feinberg, and M. Avignon, Phys. Rev. Lett. 101, 107001 (2008).
[6] R. I. Shekhter, O. Entin-Wohlman, M. Jonson, and A. Aharony, Phys. Rev. Lett. 116, 217001 (2016).
[7] R. M. Lutchyn, J. D. Sau, and S. Das Sarma, Phys. Rev. Lett. 105, 077001 (2010).
[8] Y. Oreg, G. Refael, and F. von Oppen, Phys. Rev. Lett. 105, 177002 (2010).
[9] V. Mourik, K. Zuo, S. M. Frolov, S. R. Plissard, E. P. A. M. Bakkers, and L. P. Kouwenhoven, Science 336, 1003 (2012).
[10] M. T. Deng, C. L. Yu, G. Y. Huang, M. Larsson, P. Caro , and H. Q. Xu, Nano Lett. 12, 6414 (2012).
[11] A. Das, Y. Ronen, Y. Most, Y. Oreg, M. Heiblum, and H. Shtrikman, Nat. Phys. 8, 887 (2012).
[12] H. O. H. Churchill,V. Fatemi,K.Grove-Rasmussen, M. T. Deng, P. Caro , H. Q. Xu, and C. M. Marcus, Phys. Rev. B 87, 241401 (2013).